\documentclass[11pt]{article}
\pdfoutput=1
\usepackage{authblk}
\usepackage{palatino}
\usepackage{xcolor}
\usepackage{url}
\usepackage{subfigure}
\usepackage{amssymb}
\usepackage{amsmath}
\usepackage{graphicx}
\usepackage{boxedminipage}
\usepackage{xy}
\usepackage{url}

\usepackage{ifthen}
\usepackage{algorithm}
\usepackage{algpseudocode}

\xyoption{all}

\newtheorem{theorem}{Theorem}[section]
\newtheorem{lemma}[theorem]{Lemma}

\newtheorem{proposition}{Proposition}[section]

\def\eod{\vrule height 6pt width 5pt depth 0pt}
\newenvironment{proof}{\noindent {\bf Proof:} \hspace{.4em}}
                      {\hspace*{\fill}{\eod}}
 
\newcommand{\mypara}[1] {\subsubsection*{#1}}
\newcommand{\mysmallpara}[1] {\paragraph{#1:}}
\newcommand{\eat}[1] {}
\newcommand{\att} {{\tt att}}
\newcommand{\rtot} {r_{\rm tot}}
\newcommand{\anc} {{\tt anchor}}

\newcommand{\myactive} {{\tt active}}
\newcommand{\capa} {{\tt cap}}
\newcommand{\dmax} {d_{\max}}
\newcommand{\calA} {{\cal A}}
\newcommand{\calC} {{\cal C}}
\newcommand{\calV} {{\cal V}}
\newcommand{\load} {{\tt load}}
\newcommand{\push} {{\tt pushable}}
\newcommand{\argmax} {{\rm argmax}}

\newcommand{\myin} {{\rm in}}

\newcommand{\sigmain} {\sigma_{\myin}}
\newcommand{\sigmaout} {\sigma_{\out}}
\newcommand{\pair}[2] {\langle #1, #2\rangle}
\newcommand{\wh}[1] {\widehat{#1}}
\newcommand{\xin} {x_{\myin}}
\newcommand{\yin} {y_{\myin}}

\newcommand{\myred} {{\tt Red}}
\newcommand{\myblue} {{\tt Blue}}
\newcommand{\mybrown} {{\tt Brown}}
\newcommand{\myreg} {{\tt Region}}
\newcommand{\amnt} {{\tt amnt}}
\newcommand{\rem} {{\tt rem}}
\newcommand{\wmin} {w_{\min}}

\newcommand{\out} {{\rm out}}
\newcommand{\cost} {{\rm cost}}
\newcommand{\PO} {{\sf PO}}

\newcommand{\calT} {{\cal T}}
\newcommand{\calB} {{\cal B}}
\newcommand{\calR} {{\cal R}}
\newcommand{\calBbar} {\overline{\cal B}}

\newcommand{\RP} {{\sf RP}}

\newcommand{\CSC} {{\sf CSC}}
\newcommand{\qed} {\hfill$\Box$}

\begin{document}
\title{Replica Placement on Bounded Treewidth Graphs\thanks{An abridged version of this paper is to appear in the proceedings of WADS'17.}}
\author[1]{Anshul Aggarwal}
\author[2]{Venkatesan T. Chakaravarthy}
\author[1]{Neelima Gupta}
\author[2]{\\Yogish Sabharwal}
\author[1]{Sachin Sharma}
\author[1]{Sonika Thakral\thanks{Corresponding author.}}

\affil[1]{
	University of Delhi, India.\\
	\it{ngupta@cs.du.ac.in, sonika.ta@gmail.com}
}
\affil[2]{
	IBM Research, India.\\
	\it{\{vechakra, ysabharwal\}@in.ibm.com}
}
\maketitle

\maketitle          

\begin{abstract}
We consider the replica placement problem:
given a graph and a set of clients, place replicas on a minimum set of nodes to serve all the clients; 
each client is associated with a request and maximum distance that it can travel to get 
served; there is a maximum limit (capacity)
on the amount of request a replica can serve. 
The problem falls under the general framework of capacitated set cover.
It admits an $O(\log n)$-approximation and it is 
NP-hard to approximate within a factor of $o(\log n)$.
We study the problem in terms of the treewidth $t$ of the graph 
and present an $O(t)$-approximation algorithm.
\end{abstract}

\section{Introduction}
We study a form of capacitated set cover problem \cite{Chuzhoy} called  {\em replica placement} ($\RP$) that finds
applications in settings such as data distribution by internet service providers (ISPs) and 
video on demand service delivery (e.g., \cite{1,4}).
In this problem, we are given a graph representing a network of servers and a set of clients.
The clients are connected to the network by attaching each client to a specific server.
The clients need access to a database. 
We wish to serve the clients by placing replicas (copies) of the database on a selected set of servers and clients.
While the selected clients get served by the dedicated replicas (i.e., cached copies) placed on themselves,
we serve the other clients by assigning them to the replicas on the servers.
The assignments must be done taking into account Quality of Service (QoS) and capacity constraints.
The QoS constraint stipulates a maximum distance between each client and the replica serving it.
The clients may have different demands (the volume of database requests they make)
and the capacity constraint specifies the maximum demand that a replica can handle.
The objective is to minimize the number of replicas opened.
The problem can be formally defined as follows.

\mypara{Problem Definition ($\RP$)}
The input consists of a graph $G=(\calV, E)$, a set of clients $\calA$ and a capacity $W$.
Each client $a$ is attached to a node $u\in \calV$, denoted $\att(a)$.
For each client $a\in \calA$, the input specifies a request $r(a)$ and a distance $\dmax(a)$.
For a client $a\in \calA$ and a node $u\in \calV$, let $d(a, u)$ denote the length of the
shortest path between $u$ and $\att(a)$, the node to which $a$ is attached - the length is measured
 by the number of edges and we take $d(a, u) = 0$, if $u = \att(a)$.
We say that a client $a\in \calA$ can {\em access} a node $u\in \calV$, if $d(a, u)$ is at most $\dmax(a)$.
A feasible solution consists of two parts:
(i) it identifies a subset of nodes $S\subseteq \calV$ where a replica is placed at each node in $S$;
(ii) for each client $a\in \calA$, it either opens a dedicated replica at $a$ itself for serving the client's request
or assigns the request to the replica at some node $u\in S$ accessible to $a$.
The solution must satisfy the constraint that for each node $u\in S$,
the sum of requests assigned to the replica at $u$ does not exceed $W$.
The cost of the solution is the number of replicas opened, i.e., cardinality of $S$ plus
the number of dedicated replicas opened at the clients.
The goal is to compute a solution of minimum cost.
In order to ensure feasibility, without loss of generality, we assume $r(a) \leq W$, $\forall \ a\in \calA$.
\qed

The $\RP$ problem falls under the framework of the capacitated set cover problem,
the generalization of the classical set cover problem wherein each set is
associated with a capacity specifying the number of elements it can cover.
The latter problem is known to have an $O(\log n)$-approximation algorithm \cite{Chuzhoy}.
Using the above result, we can derive an $O(\log n)$-approximation algorithm for the $\RP$ problem as well. 
On the other hand, we can easily reduce the classical dominating set problem to $\RP$:
given a graph representing an instance of the dominating set problem, we create a new client for each vertex and attach it to the vertex;
then, we set $\dmax(\cdot)=1$ for all the clients and $W=\infty$.
Since it is NP-hard to approximate the dominating set problem within a factor of $o(\log n)$ \cite{Feige},
by the above reduction, we get the same hardness result for the $\RP$ problem as well.

The $\RP$ problem is NP-hard even on the highly restricted special case where the graph is simply a path,
as can be seen via the following reduction from the bin packing problem.
Given $K$ bins of capacity $W$ and a set of items of sizes $s_1, s_2, \ldots, s_n$,
for each item $i$, we create a client $a$ with demand $r(a)=s_i$.
We then construct a path of nodes of length $K$ and attach all the clients to one end of the path and take $W$ to be the capacity of the nodes.

\mypara{Prior Results}
Prior work has studied a variant of the $\RP$ problem where the network is a directed acyclic graph (DAG),
and a client $a$ can access a node $u$ only if there is a directed path from $a$ to $u$ of the length at most $\dmax(a)$.
Under this setting, Benoit et al. \cite{benoit} considered the special case of rooted trees
and presented a greedy algorithm with an approximation ratio of $O(\Delta)$, where $\Delta$ is the maximum degree of the tree.
For the same problem, Arora et al. \cite{fst13} (overlapping set of authors) devised a constant factor approximation algorithm via LP rounding.

Progress has been made on generalizing the above result to the case of bounded treewidth DAGs.
Recall that treewidth \cite{tw4} is a classical parameter used for measuring how close a given graph is to being a tree
(a formal definition is included in Section \ref{sec:prelims}). 
For a DAG, the treewidth refers to the treewidth $t$ of the underlying undirected graph. 
Notice that the reduction from the bin-packing problem shows that the problem is NP-hard even for trees (i.e., $t=1$)
and rules out the possibility of designing an exact algorithm running in time $n^{O(t)}$ (say via dynamic programming)
or FPT algorithms with parameter $t$.

Arora et al. \cite{dag} made progress towards handling DAGs of bounded treewidth and designed an algorithm for the case of bounded-degree, bounded-treewidth graphs.
Their algorithm achieves an approximation ratio of $O(\Delta + t)$, where $\Delta$ is the maximum degree and $t$ is the treewidth of the DAG.
Their result also extends for networks comprising of bounded-degree bounded-treewidth subgraphs
connected in a tree like fashion. 

\mypara{Our Result and Discussion}
We study the $\RP$ problem on undirected graphs of bounded treewidth. 
Our main result is an $O(t)$-approximation algorithm running in time $O(n^c)$, where the exponent $c$ is a constant independent of the treewidth $t$.
In contrast to prior work, the approximation ratio depends only on the treewidth and is independent of other parameters such as the maximum degree.

Our algorithm is based on rounding solutions to a natural LP formulation, as in the case of prior work \cite{fst13,dag}.
However, the prior algorithms exploit the acyclic nature of the graphs 
and the bounded degree assumption to transform a given LP solution to a solution wherein each client is assigned to at most two replicas.
In other words, they reduce the problem to a capacitated vertex cover setting,
for which constant factor rounding algorithms are known \cite{Khuller-Saha}.

The above reduction does not extend to the case of general bounded treewidth graphs.
Our algorithm is based on an entirely different approach.
We introduce the notion of ``clustered solutions",
wherein the partially open nodes are grouped into clusters
and each client gets served only within a cluster.
We show how to transform a given LP solution
to a new solution in which a partially-open node participates in at most $(t+1)$ clusters.
This allows us to derive an overall approximation ratio $O(t)$.
The notion of clustered solutions may be applicable in other capacitated set cover settings as well.

\mypara{Other Related Work}
As mentioned earlier, the $\RP$ problem falls under the framework of the capacitated set cover problem $(\CSC)$,
which admits an $O(\log n)$-approximation algorithm \cite{Chuzhoy}.
Two versions of the $\CSC$ problem and its special cases have been considered:
soft capacity and hard capacity settings.
Our work falls under the more challenging hard capacity setting, wherein a set can be picked at most once.
The capacitated versions of the vertex cover problem (e.g., \cite{Khuller-Saha}) and dominating set problem (e.g., \cite{kao-hc})
have also been studied.  Our result applies to the capacitated dominating problem with uniform capacities 
and yields an $O(t)$-approximation algorithm.
The $\RP$ problem is also related to the capacitated facility location framework (e.g., \cite{facility-location})
However, a crucial difference is that $\RP$ is concerned only with whether or not a client can access a
facility, and its cost model does not include the distance between clients and facilities.

\section{Preliminaries}
\label{sec:prelims}
Here we define the notion of tree decomposition.
A tree decomposition of a graph $G = (V,E)$ is a pair $(X = \{ X_j : j \in J \}, T = (J,K))$,
where $T$ is a tree over the nodes $J$ and each node $j \in J$ is associated with 
a subset of vertices $X_j \subseteq V$ such that the following three conditions are satisfied: 
(i) each vertex belongs to at least one bag, i.e., $\bigcup_{j \in J} X_j = V$;
(ii) for every edge $(u,v) \in E$, there is a bag containing both $u$ and $v$; and
(iii) for all vertices $v \in V$, the set of nodes $\{j \in J : v \in X_j \}$
induces a subtree of $T$. The subsets $X_j$ are called {\em bags}.
The {\em width} of the tree decomposition is defined to be 
$\max_{j \in J} \left( |X_j|-1 \right)$.
The treewidth $t$ of a graph $G$ is the minimum width over all tree decompositions of $G$.
It is NP-hard to find the tree decomposition of minimum width, 
but fixed parameter tractable algorithms are known.

\section{Overview of the Algorithm}
Our $O(t)$-approximation algorithm is based on rounding solution to a natural LP formulation.
In this section, we present an outline of the algorithm 
highlighting its main features, deferring a detailed description to subsequent sections.
We assume that the input includes a decomposition $\calT$ of treewidth $t$ of the input network $G=(\calV, E)$.

\mypara{LP Formulation}
For each node $u\in \calV$, we introduce a variable $y(u)$ to represent the extent to which a replica is opened at $u$
and similarly, for each client $a\in \calA$, we add a variable $y(a)$ to represent the extent to which a dedicated replica is opened
at $a$ itself.
For each client $a\in \calA$ and each node $u\in \calV$ accessible to $a$,
we use a variable $x(a,u)$ to represent the extent to which $a$ is assigned to $u$.
For a client $a \in \calA$ and a node $u\in \calV$, we use the shorthand ``$a\sim u$'' to mean that $a$ can access $u$.
\begin{eqnarray}
\nonumber
\min & & \sum_{a\in \calA} y(a) \quad + \quad \sum_{u \in \calV} y(u) \\
y(a) + \sum_{u\in\calV~:~a\sim u} x(a,u) & \geq & 1 \quad \quad \quad \mbox{~~~~for all $a\in \calA$} \\
\sum_{a\in\calA~:~a\sim u} x(a,u)\cdot r(a) & \leq & y(u) \cdot W \quad \mbox{~~for all $u\in \calV$}\\
\label{const:3}
x(a,u) &\leq& y(u) \quad \quad \mbox{for all $a\in \calA$ and $u\in \calV$ with $a\sim u$}\\
0 \ \ \leq \ \ y(u), y(a) &\leq& 1 \quad \quad \quad \quad \mbox{~~~~for all $u\in \calV$ and $a\in \calA$}
\end{eqnarray}
Constraint (\ref{const:3}) stipulates that a client $a$ cannot be serviced at a node $u$ 
for an amount exceeding the extent to which $u$ is open.
For an LP solution $\sigma=\pair{x}{y}$, let $\cost(\sigma)$ denote the objective value of $\sigma$.

The following simple notations will be useful in our discussion.
With respect to an LP solution $\sigma$, we classify the nodes into three categories
based on the extent to which they are open. A node $u$ is said to be {\em fully-open}, if $y(u)=1$;
{\em partially-open}, if $0 < y(u) < 1$; and {\em fully-closed}, if $y(u) = 0$.
A client $a$ is said to be {\em assigned} to a node $u$, if $x(a,u) > 0$.
For a set of nodes $U$, let $y(U)$ denote the extent to which the vertices in $U$ are open,
i.e., $y(U) = \sum_{u\in U} y(u)$.

\mypara{Outline}
The major part of the rounding procedure involves transforming a
given LP solution $\sigmain=\pair{\xin}{\yin}$ into an {\em integrally open solution}:
wherein which each node $u\in \calV$ is either fully open or closed.
Such a solution differs from an integral solution as
a client may be assigned to multiple nodes (possibly to its own dedicated replica as well).
We address the issue easily via a cycle cancellation procedure to get
an integral solution.

The procedure for obtaining an integrally open solution works in two stages.
First it transforms the input solution into a ``clustered'' solution,
which is then transformed into an integrally open solution.
The notion of clustered solution lies at the heart of the rounding algorithm.
Intuitively, in a clustered solution, the set of partially open (and closed) nodes are partitioned into a collection of clusters $\calC$
and the clients can be partitioned into a set of corresponding groups satisfying three useful properties, as discussed below.

\begin{figure}[t]
\centering
\includegraphics[width=140mm]{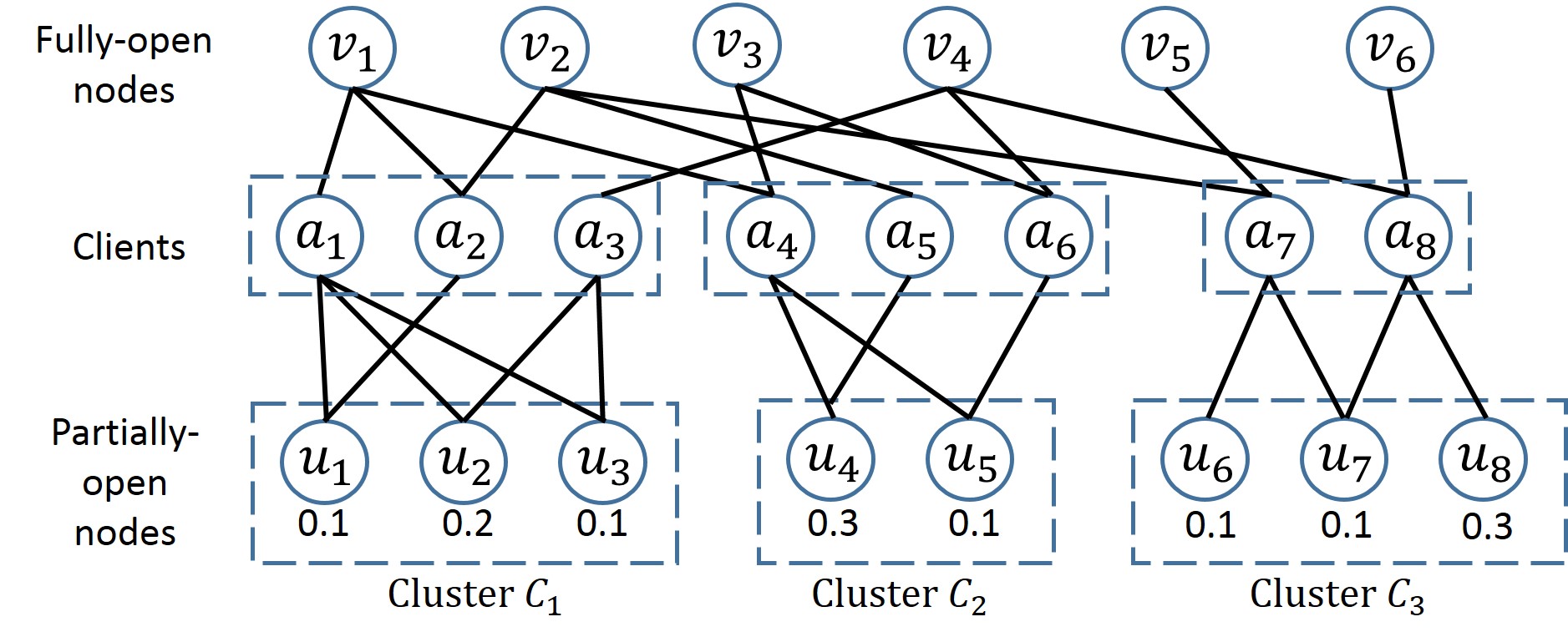}
\caption{Illustration for clustered solution. 
	Three clusters are shown $C_1, C_2$ and $C_3$, open to an extent of $0.4$, $0.4$ and $0.5$;
	the clusters are linked to the sets of fully-open nodes $\{v_1, v_2, v_4\}$, $\{v_1, v_2, v_3, v_4\}$, and $\{v_2, v_4, v_5, v_6\}$.
	The solution is $(0.5, 4)$-clustered.
}
\label{fig:clustering}
\end{figure}

Let $\sigma =\pair{x}{y}$ be an LP solution. 
It will be convenient to express the three properties using the notion of linkage:
we say that a node $u$ is {\em linked} to a node $v$, if there exists a client $a$ assigned to both $u$ and $v$.
For constants $\alpha$ and $\ell$, the solution $\sigma$ is said to be {\em $(\alpha, \ell)$-clustered},
if the set of partially-open nodes can be partitioned into a collection of clusters,
$\calC =\{C_1, C_2, \ldots, C_k\}$ (for some $k$),
such that the the following properties are true:
\begin{itemize}
	\item {\it Localization: }assignments from clients to the partially-open nodes is localized, i.e., two partially-open nodes are linked only if they belong to the same cluster.
	\item {\it Distributivity: }assignments from the clients to fully-open nodes are restricted, i.e., for any $C_j$, there are at most $\ell$ fully-open nodes that are linked to the nodes in $C_j$.
	\item {\it Bounded opening: }clusters are tiny, i.e., the total extent to which any cluster is open is at most $\alpha$, i.e., $y(C_j) < \alpha$.
\end{itemize}
Figure \ref{fig:clustering} provides an illustration.
In the first stage of the rounding algorithm, we transform the input solution $\sigmain$ into an $(\alpha, t+1)$-clustered solution
with the additional guarantee that the number of clusters is at most a constant factor of $\cost(\sigmain)$,
where $\alpha\in [0,1/2]$ is a tunable parameter. The lemma below specifies the transformation performed by the first stage. 

\begin{lemma}
	\label{lem:clustering}
	Fix any constant $\alpha \leq 1/2$. Any LP solution $\sigma$ can be transformed into 
	a $(\alpha, t+1)$-clustered solution $\sigma'$ such that
	$\cost(\sigma')$ is at most $2 + 6(t+1)\cost(\sigma)/\alpha$.
	Furthermore, the number of clusters is at most $3+ 8\cdot \cost(\sigma)/\alpha$.
\end{lemma}

At a high level, the lemma is proved by considering the tree decomposition $\calT$ of the input graph $G=(\calV, E)$
and performing a bottom-up traversal that identifies a suitable set of boundary bags. 
We use these boundary bags to split the tree into a set of disjoint regions
and create one cluster per region. We then fully open the nodes in the boundary bags
and transfer assignments from the nodes that stay partially-open to these fully-open nodes.
The transfer of assignments is performed in such a manner that clusters get localized and have distributivity of $(t+1)$. 
By carefully selecting the boundary bags, we shall enforce that each cluster is open to an extent of only $\alpha$
and that the number of clusters is also bounded. The proof is discussed in Section \ref{sec:clustering}.

The goal of the second stage is to transform a $(1/4, t+1)$-clustered solution (obtained from Lemma~\ref{lem:clustering}) 
into an integrally open solution.  
At a high level, the localization property allows us to independently 
process each cluster $C\in \calC$ and its corresponding group of clients $A$. 
The clients in $A$ are assigned to a set of fully-open nodes, say $F$.
For each node $u\in F$, we identify a suitable node $v\in C$ called the ``consort'' of $u \in C$ and fully open $v$. 
Then the idea is to transfer assignments from the non-consort nodes to the nodes in $F$ and their consorts
in such a manner that at the end, no client is assigned to the non-consort nodes.
This allows us to fully close the non-consort nodes. The localization and bounded opening properties
facilitate the above maneuver. On the other hand, the distributivity property ensures
that $F$ is at most $(t+1)$. This means that we fully open at most $(t+1)$ consorts per cluster.
Thus, overall increase in cost is at most $(t+1)|\calC|$. Since $|\calC|$ is 
guaranteed to be linear in $\cost(\sigmain)$, we get an $O(t)$ approximation factor.

\begin{lemma}
	\label{lem:part2}
	Let $\sigma = \pair{x}{y}$ be a $(1/4,t+1)$-clustered solution via a collection of clusters $\calC$.
	The solution can be transformed into an integrally open solution $\sigma' =\pair{x'}{y'}$ 
	such that $\cost(\sigma') \leq 2\cdot \cost(\sigma) + 2(t+1)|\calC|$.
\end{lemma}

Once we obtain an integrally open solution, it can easily be transformed to an integral solution
by applying a cycle cancellation strategy, as given by the following lemma. 

\begin{lemma}
	\label{lem:integral}
	Any integrally open solution $\sigma = \pair{x}{y}$ can be transformed to an integral solution $\sigma' = \pair{x'}{y'}$
	such that $\cost(\sigma') \leq 4\cdot \cost(\sigma)$.
\end{lemma}

We can transform any input LP solution $\sigmain$ into an integral solution $\sigmaout$ by applying the above three transformations
leading to the following main result of the paper. 

\begin{theorem}
\label{thm:mainthm}
The $\RP$ problem admits an $O(t)$-approximation poly-time algorithm.
\end{theorem}
\proof
We fix $\alpha = 1/4$ and apply Lemma \ref{lem:clustering} to obtain a 
solution $\sigma_1$, which is $(1/4, t+1)$-clustered via a collection of clusters $\calC$.
It is guaranteed that $\cost(\sigma_1) \leq 2 + 24(t+1)\cost(\sigmain)$ and $|\calC|\leq 3+ 32\cdot \cost(\sigmain)$.
We next apply Lemma \ref{lem:part2} on the solution $\sigma_1$ and obtain an integrally open solution
$\sigma_2$ such that $\cost(\sigma_2) \leq 2\cdot \cost(\sigma_1) + 2(t+1)|\calC|$.
Finally, we transform $\sigma_2$ into integral solution $\sigmaout$ using Lemma \ref{lem:integral}
such that $\cost(\sigmaout) \leq 4\cdot \cost(\sigma_2)$.
It follows that $\cost(\sigmaout)$ is at most $16 + 24(t+1) + 448(t+1)\cost(\sigmain)$.
Thus, the overall approximation ratio is $O(t)$. 
\qed

The constant factor involved in the approximation ratio can be improved by more careful book keeping -
however, we refrain from doing so, for the ease of exposition.
The rest of the paper is devoted to proving Lemma \ref{lem:clustering}, \ref{lem:part2} and \ref{lem:integral}.

\section{Clustered Solution: Proof of Lemma \ref{lem:clustering}}
\label{sec:clustering}
The goal is to transform a given solution into an $(\alpha, t+1)$-clustered solution with the properties claimed in the lemma.
The idea is to select a set of partially-open or closed nodes and open them fully, and then transfer assignments
from the other partially-open nodes to them in such a manner that the partially-open nodes get partitioned into clusters
satisfying the three properties of clustered solutions.
An issue in executing the above plan is that 
the capacity at a newly opened node may be exceeded during the transfer.
We circumvent the issue by first performing a pre-processing step called de-capacitation.

\subsection{De-capacitation}
Consider an LP solution $\sigma=\pair{x}{y}$ and let $u$ be a partially-open or closed node. 
The clients that can access $u$ might have been assigned to other partially-open nodes under $\sigma$.
We call the node $u$ {\em de-capacitated}, if even when all the above
assignments are transferred to $u$, the capacity at $u$ is not exceeded; meaning,
\begin{eqnarray*}
	\sum_{a\sim u}\quad \sum_{v:~a\sim v ~ \wedge ~ v\in \PO} x(a,v) &<& W,
\end{eqnarray*}
where $\PO$ is the set of partially-open nodes under $\sigma$ (including $u$).
The solution $\sigma$ is said to be {\em de-capacitated}, if all the partially-open and the closed nodes
are de-capacitated. 

\begin{figure}[t]
\begin{center}
\begin{boxedminipage}{\hsize}
\begin{tabbing}
xx\=xx\=xx\=xx\=xxx\=xxx\=\kill
\> For each partially-open node $v$ (considered in an arbitrary order)\\
\> \> For each client $a$ that can access both $u$ and $v$ (considered in an arbitrary order)\\
\> \> \> Compute capacity available at $u$: $\capa(u) = W - \sum_{b\in \calA~:~b\sim u} x(b, u) \cdot r(b)$\\
\> \> \> If $\capa(u) = 0$ exit the procedure\\
\> \> \> $\delta = \min\left\{x(a, v), \frac{\capa(u)}{r(a)}\right\}$\\
\> \> \> Increment $x(a, u)$ by $\delta$ and decrement $x(a, v)$ by $\delta$.
\end{tabbing}
\end{boxedminipage}
\end{center}
\caption{Pulling procedure for a given partially-open or closed node $u$.}
\label{fig:pulling}
\end{figure}

The preprocessing step transforms the input solution into a 
de-capacitated solution by performing a pulling procedure on the partially-open and closed nodes.
Given a partially-open or closed node $u$, the procedure transfers assignments from other partially-open nodes
to $u$, as long as the capacity at $u$ is not violated. The procedure is shown in Figure \ref{fig:pulling},
which we make use of in other components of the algorithm as well.

\begin{lemma}
\label{lem:de-capacitation}
Any LP solution $\sigma = \pair{x}{y}$ can be transformed into a de-capacitated solution $\sigma'=\pair{x'}{y'}$
such that 
$\cost(\sigma') \leq 2\cdot \cost(\sigma)$. 
\end{lemma}
\proof
We consider the partially-open and closed nodes, and process them in an arbitrary order, as follows.
Let $u$ be a partially-open or closed node. Hypothetically, consider applying the pulling procedure on $u$. 
The procedure may terminate in one of two ways: 
(i) it exits mid-way because of reaching the capacity limit;
(ii) the process executes in its entirety.
In the former case, we fully open $u$ and perform the pulling procedure on $u$.
In the latter case, the node $u$ is de-capacitated
and so, we leave it as partially-open or closed, without performing the pulling procedure.
It is clear that the above method produces a de-capacitated solution $\sigma'$.
We next analyze the cost of $\sigma'$. Let $s$ be the number of partially-open or closed nodes
converted to be fully-open. Apart from these conversions, the method does not alter the cost 
and so, $\cost(\sigma')$ is at most $s+\cost(\sigma)$.
Let the total amount of requests be $\rtot=\sum_{a\in \calA}r(a)$.
The extra cost $s$ is at most $\lfloor \rtot/W \rfloor$,
since any newly opened node is filled to its capacity.
Due to the capacity constraints, the input solution $\sigma$ must also incur a cost of at least $\lfloor \rtot/W\rfloor$.
It follows that $\cost(\sigma')$ is at most $2\cdot \cost(\sigma)$.
\qed

\subsection{Clustering}
Given Lemma \ref{lem:de-capacitation}, assume that we have a de-capacitated solution $\sigma=\pair{x}{y}$.
We next discuss how to transform $\sigma$ into an $(\alpha, t+1)$-clustered solution.
The transformation would perform a bottom-up traversal of the tree decomposition and identify a set of partially-open or closed nodes.
It would then fully open them and perform the pulling procedure on these nodes.
The advantage is that the above nodes are de-capacitated and 
so, the pulling procedure would run to its entirety (without having to exit mid-way because of reaching capacity limits).
As a consequence, the linkage between the nodes gets restricted, leading to an clustered solution.
Below we first describe the transformation and then present an analysis.

\mypara{\bf Transformation}
Consider the given tree decomposition $\calT$. We select an arbitrary bag of $\calT$ and make it the root.
A bag $P$ is said to be an {\em ancestor} of a bag $Q$, if $P$ lies on the path connecting $Q$ and the root;
in this case, $Q$ is called a {\em descendant} of $P$. We consider $P$ to be both an ancestor and descendant of itself.
A node $u$ may occur in multiple bags; among these bags the one closest to the root is called the {\em anchor}
of $u$ and it is denoted $\anc(u)$.  A {\em region} in $\calT$ refers to any set of contiguous bags 
(i.e., the set of bags induce a connected sub-tree). 

In transforming $\sigma$ into a clustered solution,
we shall encounter three types of nodes and it will be convenient
to color them as red, blue and brown. To start with, 
all the fully-open nodes are colored red and the remaining nodes 
(partially-open nodes and closed nodes) are colored blue.
The idea is to carefully select a set of blue nodes, 
fully-open them and perform the pulling procedure on these nodes; these nodes are then colored brown.
Thus, while the blue nodes are partially-open or closed, the red and the brown nodes
are fully-open, with the brown and blue nodes being de-capacitated.

\begin{figure}[t]
	\centering
\includegraphics[width=80mm]{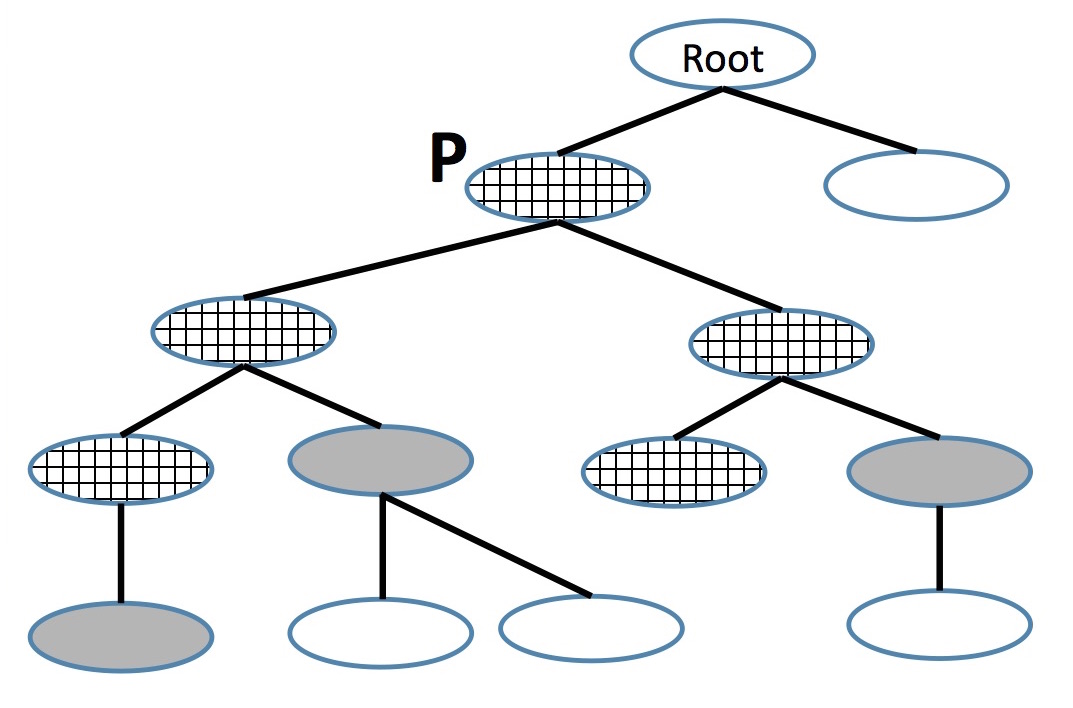}
\caption{
	Illustration for regions.
	The figure shows an example tree decomposition. The bags filled solidly represent already identified boundary bags.
	All checkered bags belong to the region headed by $P$.
}
\label{fig:region}
\end{figure}

The transformation identifies two kinds of nodes to be colored brown, {\em helpers} and {\em boundary nodes}.
We say that a red node $u\in \calV$ is {\em proper}, if it has at least one neighbor $v\in \calV$ which is a blue node.
For each such proper red node $u$, we arbitrarily select one such blue neighbor $v\in \calV$ and declare it to be the helper of $u$.
Multiple red nodes are allowed to share the same helper.
Once the helpers have been identified, we color them all brown.
The boundary brown nodes are selected via a more involved bottom-up traversal of $\calT$
that works by identifying a set $\calB$ of bags, called the {\em boundary bags}.
To start with, $\calB$ is initialized to be the empty set.
We arrange the bags in $\calT$ in any bottom-up order (i.e., a bag gets listed only after all its children are listed)
and then iteratively process each bag $P$ as per the above order.  
Consider a bag $P$. We define the {\em region headed by $P$}, denoted $\myreg(P)$, to be the set of bags $Q$ such that
$Q$ is a descendant of $P$, but not the descendant of any bag already in $\calB$.
See Figure \ref{fig:region} for an illustration. 
A blue node $u$ is said to be {\em active at $P$}, if it occurs in some bag included in $\myreg(P)$.
Let $\myactive(P)$ denote the set of blue nodes active at $P$.
We declare $P$ to be a boundary bag and add it to $\calB$ under three scenarios:
(i) $P$ is the root bag.
(ii) $P$ is the anchor of some red node.
(iii) the extent to which the nodes in $\myactive(P)$ 
are open is at least $\alpha$, i.e., $\sum_{u\in \myactive(P)} y(u) \geq \alpha$.
If $P$ is identified as a boundary bag, then we select all the blue nodes appearing in the bag and change their color to be brown.
Once the bottom-up traversal is completed, we have a set of brown nodes (helpers and boundary nodes).
We consider these nodes in any arbitrary order, open them fully, and perform the pulling procedure on them. 
We take $\sigma'$ to be the solution obtained by the above process.  This completes the construction of $\sigma'$. 
We note that a node may change its color from blue to brown in the above process, and the new color is to be considered
while determining the active sets thereafter. Notice that during the whole process of the above transformation, the solution continues to remain de-capacitated. 
A pseudocode is presented in Figure \ref{fig:clustering-pseudo}. 

\begin{figure}
\begin{center}
\begin{boxedminipage}{0.8\hsize}
\begin{tabbing}
xx\=xx\=xx\=xx\=xxx\=xxx\=\kill
\> \textbf{Input:} De-capacitated solution $\sigma = \pair{x}{y}$\\
\> \textbf{Output:} $(\alpha, t+1)$-clustered solution $\sigma' = \pair{x'}{y'}$\\
\>\\
\> $\myred \leftarrow \{u~:~\mbox{$u$ is fully-open under $\sigma$}\}$\\
\> $\myblue \leftarrow \{u~:~\mbox{$u$ is partially-open or closed under $\sigma$}\}$\\
\> $\mybrown \leftarrow \emptyset$\\
\>\\
\> // Helpers\\
\> Set helpers $H\leftarrow \emptyset$\\
\> For each node $u\in \myred$\\
\> \> if $u$ has some neighbor belonging to $\myblue$ (i.e, a proper red node) then\\
\> \> \> Let $v$ be any neighbor of $u$ belonging to $\myblue$. \\
\> \> \> Let $H \leftarrow H\cup \{v\}$\\
\> Make helpers brown: $\myblue \leftarrow \myblue - H$ and $\mybrown \leftarrow \mybrown \cup H$\\
\>\\
\>// Boundaries: Bottom-up traversal\\
\> Set $\calB \leftarrow \emptyset$\\
\> Arrange the bags in a bottom-up order\\
\> For each bag $P$ in the above order\\
\> \> if $P$ is the root, add $P$ to $\calB$\\
\> \> if $P$ is the anchor of some node $u\in \myred$, then add $P$ to $\calB$\\
\> \> $\myreg(P) \leftarrow \{Q~:~\mbox{$Q$ is a desc. of $P$, but not a desc. of any bag in $\calB$}\}$\\
\> \> $\myactive(P) \leftarrow \{u\in \myblue~:~\mbox{$u$ occurs in some bag $Q\in \myreg(P)$}\}$\\
\> \> if $\left(\sum_{u\in \myactive(P)}y(u) \geq \alpha\right)$, add $P$ to $\calB$\\
\> \> if $P$ were added to $\calB$\\
\> \> \> For each node $u\in \myblue$ occuring in $P$\\
\> \> \> \> Delete $u$ from $\myblue$ and add to $\mybrown$\\
\>\\
\>//Pulling \\
\> Arrange the nodes in $\mybrown$ in an arbitrary order\\
\> For each node $u$ in the above order\\
\> \> Fully open $u$ and perform the pulling procedure on $u$.\\
\> Output $\sigma'$ as the solution obtained above.\\
\end{tabbing}
\end{boxedminipage}
\end{center}
\caption{Pseudocode for clustering}
\label{fig:clustering-pseudo}
\end{figure}

\mypara{\bf Analysis}
We now show that $\sigma'$ is an $(\alpha, t+1)$-clustered solution. 
To start with, we have a set of red nodes that are fully-open and 
a set of blue nodes that are either partially-open or closed under $\sigma$.
The red nodes do not change color during the transformation.
On the other hand, each blue node $u$ becomes active at some boundary bag $P$.
If $u$ occurs in the bag $P$, it changes its color to brown, otherwise it stays blue.
Thus, the transformation partitions the set of originally blue nodes into a set of brown nodes and a set of nodes that stay blue.
In the following discussion, we shall use the term `blue' to refer to the nodes that stay blue.
With respect to the solution $\sigma'$, the red and brown nodes are fully-open, whereas 
the blue nodes are partially-open or closed. 

Recall that with respect to $\sigma'$, two nodes $u$ and $v$ are linked, if there is a client $a$ assigned to both $u$ and $v$.
In order to prove the properties of $(\alpha, t+1)$-clustering, we need to analyze the linkage information for the blue nodes. 
We first show that the blue nodes cannot be linked to brown nodes, by proving the following stronger observation. 

\begin{proposition}
\label{prop:swallow}
If a client $a\in \calA$ is assigned to a blue node $u$ under $\sigma'$, then $a$ cannot access any brown node $v$.
\end{proposition}
\proof
As part of the transformation, we perform the pulling procedure on the brown node $v$.
Since $\sigma$ is de-capacitated, the node $v$ is de-capacitated under $\sigma$.
As a result, the pulling procedure on $v$ would run to its entirety (without having to exit mid-way because of 
reaching the capacity limit). This means that the assignment $x(a, u)$ would get transferred to $v$.
Thus, under $\sigma'$, the client $a$ cannot remain assigned to $u$, contradicting the assumption in the lemma statement.
\qed


Proposition \ref{prop:swallow} rules out the possibility of a blue node $u$ being linked to any brown node.
Thus, $u$ may be linked to a red node or another blue node. 
The following lemmas establish a crucial property on the connectivity in these two settings.

\begin{lemma}
\label{lem:link1}
If two blue nodes $u$ and $v$ are linked under $\sigma'$,
then there must exist a path connecting $u$ and $v$ consisting of only blue nodes. 
\end{lemma}
\proof
Let $a$ be any client that is assigned to both $u$ and $v$.
Consider any shortest path $p_1$ between $u$ and $\att(a)$ (the node to which the client $a$ is attached in the network).
The path cannot contain any brown node $w$, because in this case, $d(a, w)$ would be smaller than
$d(a, u)$, making $w$ accessible to $a$. This would contradict Proposition \ref{prop:swallow}.
In a similar vein, we claim that the path cannot contain any red node.
For otherwise, traverse the path from $\att(a)$ to $u$, and let $w$ be the last red node encountered on the path.
Let $z$ be the node succeeding $w$ (it may be the case that $z=u$).
The node $z$ is blue and is a neighbor of $w$ in the graph. 
This means that $w$ is a proper red node and must have a brown helper $h$. 
We have that $d(a, w) \leq d(a, u) -1$ and $d(a, h) \leq d(a, w)+1$,
and hence $d(a, h)\leq \dmax(a)$.
This means that $a$ can access $h$, contradicting Proposition \ref{prop:swallow}.
We have shown that the path $p_1$ consists of only blue nodes. 

The same argument also shows that any shortest path $p_2$ connecting $\att(a)$ and $v$ must also consists of only blue nodes. 
The path $p_1$ connects $u$ and $\att(a)$, and the path $p_2$ connects $\att(a)$ and $v$.
By combining the two, we can construct a path $p'$ connecting $u$ and $v$.
The path $p'$ may not be simple, but we can trim it to obtain a simple path $p$ connecting $u$ and $v$.
The path $p$ contains only blue nodes. 
\qed

\begin{lemma}
\label{lem:link2}
If a blue node $u$ is linked to a red node $v$ under $\sigma'$,
then there must exist a path $p$ connecting $u$ and $v$ such that barring $v$, the path consists of only blue nodes. 
\end{lemma}
\proof
Let $a$ be a client assigned to both $u$ and $v$ under $\sigma'$.
Let $p_1$ and $p_2$ be any shortest paths connecting $\att(a)$ with $u$ and $v$, respectively.
As argued in Lemma \ref{lem:link1}, the two paths cannot contain any brown nodes
and furthermore, $p_1$ must contain only blue nodes. 
This implies that the node to which $a$ is attached, $\att(a)$, must also be a blue node.

We claim that the path $p_2$ cannot contain any red nodes, barring $v$.
By contradiction, suppose such a red node exists.
Traverse the path from $\att(a)$ to $v$.  The first node on the path is $\att(a)$, a blue node,
and continuing further, let $w$ be the first red node encountered. 
The node preceding $w$ is a blue node.
It follows that $w$ is a proper red node and so, it must have a brown helper $h$.
Furthermore, $d(a, w) \leq d(a, v)-1$ and $d(a, h) \leq d(a, w)+1$.
Thus, $d(a, h) \leq \dmax(a)$, which implies that $a$ can access the brown node $h$, contradicting Proposition \ref{prop:swallow}.

We have shown that barring $v$, the paths $p_1$ and $p_2$ consist of only blue nodes.
By combining the two paths, we can obtain  the path $p$ claimed in the lemma.
\qed

The transformation outputs a set of boundary bags $\calB$; let $\calBbar$ denote the set of non-boundary bags.
If we treat the bags in $\calB$ as cut-vertices and delete them from $\calT$,
the tree splits into a collection $\calR$ of disjoint regions.
Alternatively, these regions can be identified in the following manner.
For each bag $P\in \calB$ and each of its non-boundary child $Q\in \calBbar$,
add the region headed by $Q$ ($\myreg(Q)$) to the collection $\calR$.
Let the collection derived be $\calR=\{R_1, R_2, \ldots, R_k\}$.
It is easy to see that $\calR$ partitions $\calBbar$ and that the regions in $\calR$ are pairwise disconnected 
(not connected by edges of the tree decomposition).
We next make two observations regarding connectivity among the regions, with the second one being a generalization of the first.

\begin{proposition}
\label{prop:reg-reg1}
Consider any region $R_j\in \calR$. Let $u$ and $v$ be two nodes such that $u$ occurs only in the bags of $R_j$,
whereas $v$ does not occur in any bag of $R_j$.
Then, any path $p$ in $G$ connecting $u$ and $v$ must pass through some boundary bag $X$,
i.e., one of the nodes of $p$ must occur in $X$.
\end{proposition}

\begin{proposition}
\label{prop:reg-reg2}
Consider any region $R_j\in \calR$. Let $Q$ be the bag heading $R_j$ and let $P\in \calB$ be its parent bag.
Let $u$ and $v$ be two nodes such that $u$ occurs only in the bags of $R_j$,
$v$ does not occur in $P$ and  $\anc(v)$ does not belong to $R_j$.
Then, any path $p$ connecting $u$ and $v$ must
include a node $w\neq v$ such that $w$ occurs in $P$ or $\anc(v)$. 
\end{proposition}

The two propositions can be proved by appealing to the properties of tree decompositions.
The first follows as a direct consequence of these properties.
We can prove the second by arguing two cases:
(i) if $\anc(v)$ is a descendant of $P$, then 
the path $p$ must include a node $w\neq v$ occurring in $\anc(v)$;
(ii) if $\anc(v)$ is not a descendant of $P$,
the path must include a node $w\neq v$ occurring in $P$.

We are now ready to show that $\sigma'$ is an $(\alpha, t+1)$-clustered solution.
Towards that goal, let us suitably partition the set of partially open nodes into a collection of clusters $\calC$.  
For each region $R_j$, let $C_j$ be the set of partially open nodes that occur in some bag of $R_j$.
We take $\calC$ to be the collection $\{C_1, C_2, \ldots, C_k\}$.

Let us verify that the collection $\calC$ constructed above is indeed a partitioning of the set of partially open nodes.
Firstly, we can see that any partially open node $u$ must belong to some cluster $C_j$:
the node $u$ cannot occur in any boundary bag (for otherwise, $u$ would have turned brown)
and so, it must occur in a non-boundary bag found in some region $R_j$ and would get included in $C_j$.
Secondly, any partially open node $u$ cannot belong to two clusters $C_i$ and $C_j$.
For otherwise, $u$ must occur in some bags $Q_1\in R_i$ and $Q_2\in R_j$.
Since $R_i$ and $R_j$ are disconnected, the (unique) path connecting $Q_1$ and $Q_2$ in $\calT$
must pass through some boundary bag $P$.
By the properties of tree decomposition, the node $u$ must also occur in $P$.
In this case, $u$ would have turned brown, contradicting the assumption that $u$ is partially open, and hence blue.

We next argue that $\calC$ satisfies the three properties of localization, distributivity and
bounded opening. However, the number of clusters in the collection may exceed the bound claimed in Lemma \ref{lem:clustering}.
Later, we show that the issue can be easily rectified by suitably merging the clusters.

\begin{lemma}
\label{lem:bdd-cluster}
The solution $\sigma'$ is $(\alpha, t+1)$-clustered.
\end{lemma}
\proof
We prove the collection $\calC$ satisfies the three properties.

{\it Localization: }
We need to show that any two linked blue nodes $u$ and $v$ belong to the same cluster.
By contradiction, suppose that there exist two blue nodes $u$ and $v$ belonging to two different clusters $C_i$ and $C_j$
such that a common client $a$ is assigned to both of them under $\sigma'$.
Lemma \ref{lem:link1} shows that $u$ and $v$ are connected by a path $p$ consisting only of blue nodes.
By the construction of the clusters, $u$ and $v$ occur only in the bags of the regions $R_i$ and $R_j$, respectively.
Thus, by Proposition \ref{prop:reg-reg1}, some node $w$ found in $p$ must occur in some boundary bag $P$.
However, in this case, the transformation would have turned the blue node $w$ to a brown node,
contradicting the fact that $w$ stayed blue.

{\it Distributivity: }
Consider any cluster $C_j$ and any node $u\in C_j$. 
Let $Q$ be the bag heading the corresponding region $R_j$ and let $P$ be the parent bag of $Q$.
We claim that any red node $v$ linked to $u$ must occur in $P$.
By contradiction suppose $v$ does not occur in $P$.
By Lemma \ref{lem:link2}, there must exist a path $p$ connecting $u$ and $v$,
which is made of all blue nodes, barring $v$.
The bag $\wh{P}=\anc(v)$ cannot belong to the region $R_j$; 
for otherwise, the transformation would have made $\wh{P}$ into a boundary bag,
but the region $R_j$ consists of only non-boundary bags.
Thus, Proposition \ref{prop:reg-reg2} implies that the path $p$ must include a node $w\neq v$ such that $w$ occurs in $P$ or $\anc(v)$.
Both $P$ and $\anc(v)$ are boundary bags and $w$ is a blue node.
In this case, the transformation would have turned $w$ to a brown node,
contradicting the fact that $w$ stayed blue.
The claim implies that all the red nodes that are linked to the blue nodes in $C_j$ occur in the bag $P$.
Since $\calT$ is a decomposition of width $t$, $P$ can contain at most $t+1$ elements.
Thus, the blue nodes in $C_j$ can be linked to at most $t+1$ red nodes.
In the solution $\sigma'$, the red and brown nodes are fully-open. 
By Proposition \ref{prop:swallow}, the brown nodes cannot be linked to blue nodes.
We have thus proved that the clustering has distributivity parameter $t+1$.

{\it Bounded opening: }
We claim that each cluster $C_j$ is open to an extent of less than $\alpha$, i.e., $y(C_j) < \alpha$.
For otherwise, consider the corresponding region $R_j$ and the bag $Q$ heading $R_j$.
Notice that if $y(C_j) \geq \alpha$, the transformation would have made $Q$ itself to a boundary bag,
but any region in the collection $\calR$ contains only non-boundary nodes.
\qed

\mysmallpara{Cost Analysis}
Here we analyze the solution $\sigma'=\pair{x'}{y'}$ and prove the bound claimed in Lemma \ref{lem:clustering}.
Let $\myred$, $\myblue$ and $\mybrown$ denote the set of red, blue and brown nodes.
Then, $\cost(\sigma')$ is given by $|\myred|+|\mybrown| + y'(\myblue) + y(\calA)$,
where $y(\calA)$ represents the extent to which dedicated replicas are opened, 
i.e., $y(\calA) = \sum_{a\in \calA} y(a)$.
The red nodes do not change their color, for any blue node $u$,
the extent to which it is open does not change and similarly, for any client $a$, $y(a)$ does not change.
Thus, $|\myred| + y'(\myblue) + y'(\calA) \leq \cost(\sigma)$
and hence, $\cost(\sigma') \leq \cost(\sigma) + |\mybrown|$.
We create a brown helper node for each red node.
Furthermore, for each boundary bag $P\in \calB$, we convert all the blue nodes in $P$ to be brown,
and the number of such blue nodes is at most $(t+1)$. 
Thus,  $|\mybrown| \leq |\myred| + (t+1)|\calB|$. 
A bag $P$ is made a boundary bag under one of the three scenarios. 
(i) $P$ is the root bag; (ii) $P$ is the anchor of some red node;
(iii) the total extent to which the nodes in $\myactive(P)$ are open is at least $\alpha$.
The number of boundary bags of the first two types are $1+|\myred|$.
Regarding the third scenario, notice that each originally blue node becomes active at a unique boundary bag.
This is because, each originally blue node becomes active at some boundary piece $P$.
If it occurs in $P$, then it turns brown and otherwise, by the properties of tree decomposition
it cannot occur in the region of any other boundary bag.
The total extent to which these originally blue nodes are open is at most $\cost(\sigma)$.
Thus, the number boundary bags of the third type is at most $\lceil \cost(\sigma)/\alpha \rceil$.
Therefore, 
\[
	|\calB| \leq 1 + |\myred| + \lceil\cost(\sigma)/\alpha\rceil \leq 2 + |\myred| + \cost(\sigma)/\alpha.
\]
It follows that $\cost(\sigma')$ is at most  $\cost(\sigma) + |\myred| + (t+1)(2 + |\myred| + \cost(\sigma)/\alpha)$.
A simple arithmetic shows that $\cost(\sigma')$ is at most $2+3(t+1)\cost(\sigma)/\alpha$
(we use the fact that $|\myred|\leq \cost(\sigma)$ and our assumption that 
the parameter $\alpha$ is at most $1/2$).
The preprocessing step of de-capacitation incurs a 2-factor increase in cost.
Taking this into account, we get the cost bound claimed in the statement of Lemma \ref{lem:clustering}.

\mysmallpara{Number of Clusters}
As mentioned earlier, an issue with the collection $\calC$ is that it may have more clusters than the bound claimed in Lemma \ref{lem:clustering}.
We reduce the number of clusters by suitably merging the clusters.
Consider each boundary bag $P$. 
All the non-boundary children of $P$ have a corresponding cluster in $\calC$
and let $\calC_P$ denote the collection of these clusters. 
We start with the collection $\calC_P$ and repeatedly perform the following merging operation.
Select any two clusters $C$ and $C'$ from $\calC_P$ such that 
$y(C)\leq \alpha/2$ and $y(C')\leq \alpha/2$ and merge the two into a single cluster.
The process is stopped when we cannot find two such clusters.
This way we get a set of new clusters all of which are open to an extent of at most $\alpha$.
Furthermore, except for perhaps a single cluster, all the others are open to an extent of at least $\alpha/2$;
we refer to these as {\em normal clusters} and the exceptional one as {\em abnormal}.
We perform this processing for all the boundary bags and obtain a new collection $\calC'$.
The number of abnormal clusters is at most $|\calB|$.
The collection $\calC'$ is a partitioning of $\myblue$ and each normal cluster is open to an extent of at least $\alpha/2$.
Thus, the number of normal clusters can be at most $\lceil y'(\myblue)/(\alpha/2)\rceil$,
which is at most $\lceil 2\cost(\sigma)/\alpha\rceil$.
Hence, the total number of clusters in $\calC'$ is at most $3 + 4 \cost(\sigma)/\alpha$.
The process of merging does not affect distributivity: as shown in the proof Lemma \ref{lem:bdd-cluster},
for any two merged clusters, the partially-open nodes in the clusters can only be linked to the fully-open 
nodes found in the parent boundary piece and the count of such fully-open nodes can be at most $(t+1)$.
The preprocessing step of de-capacitation increases cost by $2$-factor.
Taking this into account, we get the bound on number of clusters claimed in the statement of Lemma \ref{lem:clustering}.

\begin{figure}
\begin{center}
\begin{boxedminipage}{0.8\hsize}
\begin{tabbing}
xx\=xx\=xx\=xx\=xxx\=xxx\=\kill
Let $A$ be the set of clients assigned to nodes in $C$\\
Let $F=\{u_1, u_2, \ldots, u_{t+1}\}$ be the fully-open nodes linked to nodes in $C$\\
Apply Proposition \ref{prop:single-red} to get a solution $\sigma' =\pair{x'}{y'}$\\
For $i\leq t+1$, let $A_i\subset A$ be the set of clients assigned to $u_i$.\\
\\
/* Selection of consorts */\\
Let $L \leftarrow \emptyset$\\
For $i$ to $1$ to $t+1$\\
\> For each node $v\in C$: let $r(A_i, v) = \sum_{a\in A_i:a\sim v} r(a)$.\\
\> Let $v_i \leftarrow \argmax_{v\in C-L} r(A_i, v)$ and add $v_i$ to $L$.\\
Let $C' \leftarrow C-L$\\
\\
/* Push from nodes in $F$ to $L$\\
For $i$ from $1$ to $t+1$\\
\> Let load to push: $\push(u_i, v_i) = \sum_{a\in A_i:a\sim v_i} x(a, u_i)r(a)$\\
\> Let remaining load: $\rem \leftarrow \push(u_i, v_i)$\\
\> For each $a\in A_i$ such that $a \sim v_i$ (considered in an arbitrary order)\\
\> \> Let $\amnt \leftarrow \min\{\rem, x(a, u_i)r(a)\}$\\
\> \> Let $\delta \leftarrow \amnt/r(a)$\\
\> \> $x'(a, v_i) \leftarrow x'(a, v_i) + \delta$ and $x'(a, u_i) \leftarrow x'(a, u_i) - \delta$\\
\> \> $\rem \leftarrow \rem - \amnt$\\
\> \> If $\rem == 0$ exit loop and go to next $i$.\\
\\
/* Transfer load from $C'$ to $F$ */\\
For each node $v\in C'$ and each node $u_i\in F$\\
\> For each client $a\in A_i$ and $a\sim v$\\
\> \> $x'(a, u_i) \leftarrow x'(a, u_i) + x'(a, v)$ and $x'(a, v) \leftarrow 0$\\
\end{tabbing}
\end{boxedminipage}
\end{center}
\caption{Pseudocode for processing for a Cluster $C$}
\label{fig:C}
\end{figure}

\section{Integrally Open Solution: Proof of Lemma \ref{lem:part2}}
\label{sec:part2}
Our goal is to transform a given $(1/4, t+1)$-clustered solution $\sigma=\pair{x}{y}$ into an integrally open solution $\sigma'$.
We classify the clients into two groups, {\em small} and {\em large},
based on the extent to which they are served by dedicated replicas:
a client $a\in \calA$ said to be {\em small}, if $y(a) < 1/2$, and it is said to be {\em large} otherwise.
Let $\calA_s$ and $\calA_l$ denote the set of small and large clients, respectively.

We pre-process the solution $\sigma$ by opening a dedicated replica at each large client $a$
and removing its assignments to the nodes
(set $y(a) = 1$ and set $x(a, u) = 0$ for all nodes $u$ accessible to $a$).
We see that the transformation at most doubles the cost
and the solution remains $(1/4, t+1)$-clustered.

Consider the pre-processed solution $\sigma$. 
Let $\calC$ denote the set of clusters (of the partially-open nodes) under $\sigma$.
For each cluster $C\in \calC$, we shall fully open a selected set of at most $2(t+1)$ nodes
and fully close  rest of the nodes in it.

We now describe the processing for a cluster $C\in \calC$.
Let $A\subseteq \calA_s$ denote the set of clients assigned to the nodes in $C$. 
By the distributivity property, these clients are assigned to
at most $(t+1)$ fully-open nodes, denoted $F=\{u_1, u_2, \ldots, u_{t+1}\}$.
A client $a\in A$ may be assigned to multiple nodes from $F$.
In our procedure, it would be convenient if each client is assigned to at most one node from $F$
and we obtain such a structure using the following proposition.

The proposition is proved via a cycle cancellation procedure
that transfers assignments amongst the nodes in $F$. 
The procedure can ensure that, except for at most $|F|$ clients,  every other client $a\in \calA$ is assigned to at most one node from $F$.
We open dedicated replicas at the exceptional clients and this results in an cost increase of at most $|F|$.

\begin{proposition}
\label{prop:single-red}
Given a solution $\sigma=\pair{x}{y}$, a set of fully-open nodes $F$ and a set of clients $A$,
we can obtain a solution $\sigma'=\pair{x'}{y'}$ such that
each client $a\in A$ is assigned to at most one node from $F$. 
Furthermore, the transformation does not alter the other assignments, 
i.e., for any node $u\in \calV$ and any client $a\in \calA$, if $u\not\in F$ or $a\not\in A$, 
then $x'(a,u) = x(a,u)$. Moreover, $\cost(\sigma') \leq \cost(\sigma) + |F|$.
\end{proposition}
\proof
Construct an edge-weighted bipartite graph with nodes in $F$ on one side
and the clients in $A$ on the other side. For a pair of nodes $u\in F$ and $a\in A$,
add an edge between the two, if $a$ is assigned to $u$ under $\sigma$.
In this case, we imagine that $a$ imposes a load of $x(a, u)r(a)$ on the node $u$
and represent the above quantity as the weight on the edge.
The plan is to employ a standard cycle-cancellation strategy and make the graph acyclic.
Towards that goal, consider any cycle in the graph. Since the graph is bipartite,
the cycle must be of even length. Partition the edges in the cycles into two groups,
odd and even, by alternating on the cycle. 
Let $e=(a, u)$ be the edge having the least weight and let $\wmin = x(a, u)r(a)$.
Assume without loss of generality that $e$ is an odd edge.
The idea is to decrease the load on all the odd edges by an amount $\wmin$ and
increase the load on all the even edges by the same amount.
This can be accomplished by adjusting the assignments as follows.
For each edge $e' = (a', u')$, compute $\delta = \wmin/r(a')$.
If $e'$ is an odd edge, increase $x(a', u')$ by an amount $\delta$,
and otherwise, decrease $x(a', u')$ by an amount $\delta$.
The edge weights are recomputed accordingly.
The above process makes the assignment $x(a, u)$ to be zero
and so, we can delete the edge, thereby breaking the cycle.
We repeat the process until the bipartite graph becomes acyclic, i.e., a forest.

Consider the resultant LP solution. The forest provides us information on the nodes that the clients are assigned to:
a client $a\in A$ is assigned to a node $u\in F$, if  $u$ is a neighbor of $a$ in the forest.
Thus, any client $a\in A$ appearing as a leaf (vertex of degree one)
is assigned to only a single node from $F$. These clients satisfy the property claimed in the proposition.
This leaves us with having to deal with clients having multiple neighbors
-- let $A'$ denote the set of such clients.
We handle these clients simply by opening a dedicated replica at the client node itself.
The process produces a solution $\sigma'$ wherein each client $a\in A$ is assigned to at most one node $u\in F$.

The above process incurs an extra cost of one unit per dedicated replica and so, the total increase in cost is $|A'|$.  
It is not difficult to argue that $|A'| \leq |F|$. To prove this, we shall produce a one-to-one mapping from $A'$ to $F$.
Consider each tree in the forest and root it at an arbitrary node from $F$. 
Since the graph is bipartite, the nodes from $F$ and the clients from $A$ appear in alternate levels of the tree.
Thus, for any client $a\in A'$, all its children are from $F$.
For each client $a\in A'$, pick one of its children $u\in F$ and map $a$ to $u$.
This is a one-to-one mapping and so, $|A'|\leq |F|$. We have shown that $\cost(\sigma') \leq \cost(\sigma) + |F|$.
\qed

%


The proposition does not alter the other assignments and so, its output solution is also $(1/4, t+1)$-clustered.
Given the proposition and the pre-processing, we can assume that $\sigma=\pair{x}{y}$ is $(1/4, t+1)$-clustered 
wherein each client $a\in A$ is assigned to at most one node from $F$ and that $y(a) < 1/2$.
For each node $u_i\in F$, let $A_i\subseteq A$ denote the set of clients assigned to the node $u_i$.
The proposition guarantees that these sets are disjoint. 

For a node $v$ and a client $a$, let $\load(a, v)$ denote the amount of load imposed by $a$ on $v$ towards the capacity:
$\load(a, v) = x(a, v)r(a)$. It will be convenient to define the notion over sets of clients and nodes.
For a set of clients $B$ and a set of nodes $U$, let $\load(B, U)$ denote the load imposed by the clients
in $B$ on the nodes $U$: $\load(B, U) =\sum_{a\in B, v\in U:a\sim v}x(a,v)r(a)$;
when the sets are singletons, we shall omit the curly braces.
Similarly, for a subset $C'\subseteq C$, let $\load(C') = \sum_{v\in C'} \load(v)$.

The intuition behind the remaining transformation is as follows.
We shall identify a suitable set of nodes $L=\{v_1, v_2, \ldots, v_{t+1}\}$ from $C$,
with $v_i$ being called the {\em consort} of $u_i \in C$, and fully open all these nodes.
Then, we consider the non-consort nodes $C'=C-L$ and for each $i\leq t+1$, we transfer the load $\load(A_i, C')$ to the node $u_i$.
As a result, no clients are assigned to the non-consort nodes any more and so, they can be fully closed.
In order to execute the transfer, for each $i\leq t+1$, we create space in $u_i$ by pushing a load equivalent to $\load(A_i, C')$ 
from $u_i$ to its (fully-opened) consort $v_i$.
The  amount of load $\load(A_i, C')$ involved in the transfer is very small:
the bounded opening property ensures that $y(C) < 1/4$ and thus, $\load(A_i, C') < W/4$.
The fully-opened consort $v_i$ has enough additional space to receive the load:
$y(v_i)\leq 1/4$ and so, $\load(A, v_i) < W/4$, which means that if we fully open the consort, 
we get an additional space of $(3/4)W$.
However, an important issue is that a consort $v_i$ may not be accessible to all the clients in $A_i$.
Therefore, we need to carefully choose the consorts in such a manner that each 
fully open node $u_i$ has enough load accessible to the consort $v_i$
that can be pushed to $v_i$. 
Towards this purpose, we define the notion of {\em pushable load}.
For a node $u_i\in F$ and a node $v\in C$, let $\push(u_i, v)$ denote the amount of load on $u_i$ that is accessible to 
$v$:  $\push(u_i, v) = \sum_{a\in A_i:a\sim v} x(a, u_i)r(a)$. We next show how to identify a suitable set of consorts
such that the pushable load is more than the load that we wish to transfer. 

\begin{lemma}
\label{lem:io-AAA}
We can find a set of nodes $L=\{v_1, v_2, \ldots, v_{t+1}\}$ such that for all $i\leq t+1$, $\push(u_i, v) \geq \load(A_i, C')$. 
\end{lemma}
\proof
For a set of clients $B$, let $r(B)$ denote the sum of requests of the clients in $B$. For a node $v$, let $r(B,v)$ denote the sum of requests of the clients in $B$ that can access $v$, i.e.,
$r(B, v) = \sum_{a\in B:a\sim v} r(a)$.

We identify the required set via a greedy procedure. 
Initialize $L=\emptyset$ and iterate over the nodes $u_1, u_2, \ldots, u_{t+1}$.
For each node $u_i$, select $v_i=\argmax_{v\in C-L} r(A_i, v)$
and add $v_i$ to $L$. 

Let $L$ be the set identified by the above procedure and let $C' = C-L$.
Fix any $i\leq t+1$.  We derive a bound on $\load(A_i, C')$:
\begin{eqnarray*}
	\load(A_i, C') 
	&=& \sum_{v\in C'} \quad \sum_{a\in A_i:a\sim v} x(a, v) r(a) \quad \leq \quad \sum_{v\in C'} y(v) \sum_{a\in A_i:a\sim v} r(a)\\
	&\leq& \sum_{v\in C'} y(v) r(A_i, v) \quad \leq \quad r(A_i, v_i)\sum_{v\in C'} y(v)\quad < \quad (1/4)r(A_i, v_i)
\end{eqnarray*}
The second statement follows from the LP constraint (\ref{const:3}),
whereas the third statement is by the definition of $r(A_i, v)$.
The fourth statement follows from the construction
and the last statement follows from the bounded opening property.

For any client $a\in A_i$, the solution has opened a dedicated replica to an extent of $y(a)$
and the remaining assignment of $1-y(a)$ is going to the nodes.
Our construction has ensured that $a$ is a small client and so $y(a) < 1/2$.
This means that the client $a$ is assigned to an extent of at least $1/2$ to the nodes in the cluster.
Furthermore, the only nodes to which the client is assigned are $u_i$ and the nodes in the cluster $C$. 
Since $y(C) < 1/4$, the total extent to which the client $a$ is assigned to the nodes in $C$
is less than $1/4$. This implies that $x(a, u_i) \geq 1/4$. 
Therefore, 
\[
\push(u_i, v_i) \quad = \quad \sum_{a\in A_i:a\sim v_i} x(a, u_i)r(a) \quad \geq \quad (1/4)\sum_{a\in A_i:a \sim v_i} r(a) \quad = \quad (1/4)r(A_i, v_i).
\]
We have proved the lemma.
\qed

We have shown that each node $u_i$ has a load of at least $\load(A_i, C')$ which can be pushed to its consort $v_i$.
As observed earlier $\load(A_i, C') < W/4$ and $\load(A_i, v_i) < W/4$.
Hence, when we fully open the consort, we get an additional space of $(3/4)W$, which is sufficient to receive the load from $u_i$.
The pseudo-code for processing a cluster $C$ is shown in Figure \ref{fig:C}.

Given the above discussion, we iteratively consider each cluster $C_j\in \calC$ and perform the above transformation.
This results in $(t+1)$ consorts from $C_j$ being fully-opened and all the other nodes in $C_j$
being fully closed. At the end of processing all the clusters, we get a solution in which each node
either fully open or fully close. 
For each cluster $C_j$, we incur an extra cost of at most $(t+1)$ while applying Proposition \ref{prop:single-red}, 
and an additional cost of $(t+1)$ for opening the consorts.
Thus, the cost increases by at most $2(t+1)|\calC|$.

\section{Integral Solution: Proof of Lemma \ref{lem:integral}}
\label{sec:integral}
An integrally open solution falls short from being an integral solution in two aspects:
(i) a client may be assigned to more than one node;
(ii) a client may be served partly by a dedicated replica and partly by the network nodes.
We address the first issue by appealing to Proposition \ref{prop:single-red} via taking
$F$ to be the set of all fully open nodes and $A$ to be the problematic clients.
In the resultant solution each client is assigned to at most one fully open node 
and the cost can increase by a factor of at most two.
The second issue is addressed by the following proposition.

\begin{proposition}
Let $\sigma=\pair{x}{y}$ be an integrally open solution in which 
each client is assigned to at most one node. It can be transformed into an integral solution $\sigma'$
such that $\cost(\sigma') \leq 2\cdot \cost(\sigma)$.
\end{proposition}
\proof
We iteratively consider each full open node $u$. Let $A\subseteq \calA$ denote the clients assigned to $u$.
Each client $a\in A$ is served partly by its own dedicated replica to an extent of $y(a)$, while
the remaining request of $1-y(a)$ is assigned to $u$. We wish to obtain a solution wherein each 
at most one client is assigned to $u$. Suppose multiple clients are assigned to $u$.
Choose any two such clients $a$ and $b$. Without loss of generality, assume that $r(a) \geq r(b)$.
Let $\delta = \min\{x(a, u), y(b)\}$. Decrease $x(a, u)$ and $y(b)$ by $\delta$, 
and increase $y(a)$ and $x(b, u)$ by $\delta$. The assumption that $r(a) \geq r(b)$ ensures that
the above transfers do not violate the capacity constraint at the node $u$.
The transfer results in either $a$ or $b$ getting fully served by a dedicated replica 
and no longer being assigned to $u$. By repeating the process, we can derive a solution 
wherein at most one client $a$ is assigned to $u$. We then open a dedicated replica at the 
specified node $a$ and remove its assignment to $u$ (set $y(a) = 1$ and $x(a, u) = 0$).
The procedure is repeated for all fully open nodes, leading to an integral solution $\sigma'$.
The cost increases by at most one unit for each full open node. Thus, $\cost(\sigma') \leq 2\cdot \cost(\sigma)$.
\qed

We convert the input integrally open solution $\sigma$ in to an integral solution
$\sigma'$ by applying the above two steps.
Each step incurs a $2$-factor increase in cost and thus, $\cost(\sigma')$ is at  most $4\cdot \cost(\sigma)$.

\bibliographystyle{plain}
\bibliography{main}

\begin{thebibliography}{10}

\bibitem{dag}
S.~Arora, V.~Chakaravarthy, K.~Gupta, N.~Gupta, and Y.~Sabharwal.
\newblock Replica placement on directed acyclic graphs.
\newblock In V.~Raman and S.~Suresh, editors, {\em Proceedings of the 34th
  International Conference on Foundation of Software Technology and Theoretical
  Computer Science (FSTTCS)}, pages 213--225, 2014.

\bibitem{fst13}
S.~Arora, V.~Chakaravarthy, N.~Gupta, K.~Mukherjee, and Y.~Sabharwal.
\newblock Replica placement via capacitated vertex cover.
\newblock In A.~Seth and N.~Vishnoi, editors, {\em Proceedings of the 33rd
  International Conference on Foundations of Software Technology and
  Theoretical Computer Science (FSTTCS)}, pages 263--274, 2013.

\bibitem{benoit}
A.~Benoit, H.~Larchev{\^{e}}que, and P.~Renaud{-}Goud.
\newblock Optimal algorithms and approximation algorithms for replica placement
  with distance constraints in tree networks.
\newblock In {\em Proceedings of the 26th {IEEE} International Parallel and
  Distributed Processing Symposium (IPDPS)}, pages 1022--1033, 2012.

\bibitem{tw4}
H.~Bodlaender and A.~Koster.
\newblock Combinatorial optimization on graphs of bounded treewidth.
\newblock {\em Computer Journal}, 51(3):255--269, 2008.

\bibitem{Chuzhoy}
J.~Chuzhoy and J.~Naor.
\newblock Covering problems with hard capacities.
\newblock {\em SIAM Journal of Computing}, 36(2):498--515, 2006.

\bibitem{1}
I.~Cidon, S.~Kutten, and R.~Soffer.
\newblock Optimal allocation of electronic content.
\newblock {\em Computer Networks}, 40:205--218, 2002.

\bibitem{Feige}
U.~Feige.
\newblock A threshold of ln \emph{n} for approximating set cover.
\newblock {\em Journal of the {ACM}}, 45(4):634--652, 1998.

\bibitem{4}
K.~Kalpakis, K.~Dasgupta, and O.~Wolfson.
\newblock Optimal placement of replicas in trees with read, write, and storage
  costs.
\newblock {\em IEEE Transactions on Parallel and Distributed Systems},
  12:628--637, 2001.

\bibitem{kao-hc}
M.~Kao, H.~Chen, and D.~Lee.
\newblock Capacitated domination: Problem complexity and approximation
  algorithms.
\newblock {\em Algorithmica}, 72(1):1--43, 2015.

\bibitem{facility-location}
R.~Levi, D.~Shmoys, and C.~Swamy.
\newblock {LP}-based approximation algorithms for capacitated facility
  location.
\newblock {\em Math. Program.}, 131(1-2):365--379, 2012.

\bibitem{Khuller-Saha}
B.~Saha and S.~Khuller.
\newblock Set cover revisited: {H}ypergraph cover with hard capacities.
\newblock In A.~Czumaj, K.~Mehlhorn, A.~Pitts, and R.~Wattenhofer, editors,
  {\em Proceedings of the 39th International Colloquium on Automata, Languages,
  and Programming (ICALP)}, volume 7391 of {\em LNCS}, pages 762--773.
  Springer, 2012.

\end{thebibliography}

\end{document}